\documentclass[aps,prl,twocolumn,superscriptaddress,groupedaddress,floatfix]{revtex4}
\usepackage{amsmath}
\usepackage{amsfonts}
\usepackage{amssymb}
\usepackage{graphicx}
\usepackage{bm}

\begin{document}

\title{Interacting Anisotropic Dirac Fermions in  Strained Graphene and Related Systems}

\author{Anand Sharma} 

\author{Valeri N. Kotov}
\affiliation{Department of Physics, University of Vermont, 82 University Place, Burlington, Vermont 05405, USA}

\author{Antonio H. Castro Neto}
\affiliation{Department of Physics, Boston University, 590 Commonwealth Avenue, Boston, Massachusetts 02215, USA}
\affiliation{Graphene Research Centre and Department of Physics, National University of Singapore,
2 Science Drive 3, Singapore 117542}

\begin{abstract}
We study the role of   long-range  electron-electron interactions in a system of two-dimensional anisotropic Dirac
 fermions, which naturally  appear  in uniaxially strained graphene, graphene in external potentials, some strongly
  anisotropic topological insulators, and engineered  anisotropic graphene structures.  We find that
  while for small interactions and anisotropy the  system restores the conventional isotropic Dirac liquid
   behavior,  strong enough anisotropy can lead to the formation of a 
   quasi-one dimensional  electronic phase with dominant charge order  (anisotropic excitonic insulator). 
\end{abstract}

\date{\today}


\maketitle

\begin{figure}[b]
\centering
\includegraphics[width=0.38\textwidth]{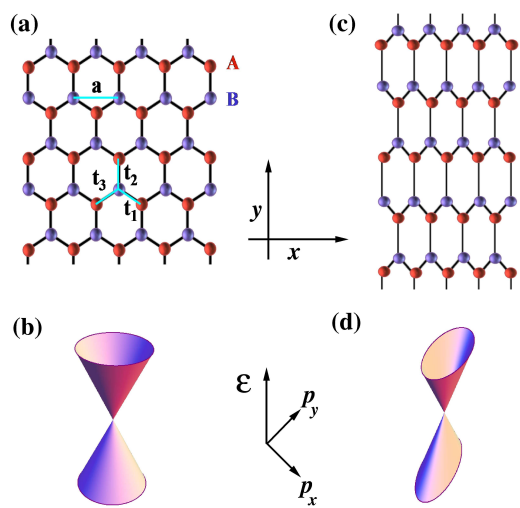}
\caption{\label{fig:1} (Color online)  Honeycomb lattice 
 for (a) undeformed  and  (c) unixially deformed  graphene along the $y$ (armchair) direction.
  The corresponding isotropic (b) and anisotropic (d)  Dirac cones are shown.} 
\end{figure}

Since the isolation of graphene  \cite{novoselov1},  a  two dimensional (2D) allotrope of carbon, there has been 
substantial theoretical and experimental effort to understand and utilize its remarkable
 mechanical \cite{lee1}, thermal \cite{balandin}, electronic \cite{castroneto} and 
 transport \cite{peres} properties.
 The charge carriers in a pristine graphene sheet move on a honeycomb lattice (Fig.~\ref{fig:1}(a)) as if they were
 massless relativistic particles. The physical properties at low energies are governed by the Dirac Hamiltonian
  resulting in a linear dispersion and isotropic  cones  (with circular cross section)  near the Dirac points  (Fig.~\ref{fig:1}(b)).
  Mechanical
   deformations, such as uniaxial strain  for example  along  the armchair direction (Fig.~\ref{fig:1}(c)), can lead to  anisotropic dispersion 
  with the formation of elliptical  Dirac cones (Fig.~\ref{fig:1}(d)) \cite{pereira1}. 
 Within the tight-binding and ab initio schemes,  the   anisotropic Dirac
   dispersion is found for weak to moderate uniaxial strain in any direction, while for
    large strain the electronic structure becomes very different along the armchair and zig-zag directions,
    ultimately leading to cone merger  and gap formation in the latter case \cite{pereira1}.
 It has also been shown that anisotropic Dirac cones can be formed by applying an external periodic
   potential \cite{park-rusponi} on graphene. 
   Such changes of the band structure can provide exciting possibilities for ``strain engineering," i.e.
   manipulation of graphene's electronic, optical, etc. properties by applying lattice deformations or potentials \cite{pereira2}.
   Strongly anisotropic Dirac cones can also  appear in certain topological insulators \cite{topo},
    and, in a recent development,  highly tunable honeycomb optical lattices and molecular graphene systems
     have been created \cite{optical},
      providing a possible route towards exploring  various anisotropic phases.
      
      The subject of the present work is the interplay between Dirac fermion anisotropy and electron interactions. 
  It is known that near the Dirac point, close to charge neutrality, unscreened long-range electron-electron interactions
   in isotropic graphene can manifest themselves in a variety of
   ways \cite{kotov}; perhaps most spectacularly interactions lead to logarithmic renormalization of the electron
    spectrum leading to reshaping of the Dirac cones (velocity increase),  observed in undeformed suspended
     graphene \cite{elias}. 
     Previously anisotropic Dirac fermions have been studied in  QED$_3$-type models,  relevant for
      the cuprate superconductors, where the fermion anisotropy was found to be irrelevant in 
       renormalization group (RG) sense \cite{vafek}, i.e. the systems flows towards the isotropic limit.
        This behavior can be reversed,  i.e. the anisotropy increases,  if the fermions couple to a nematic order parameter,  
        as suggested for the cuprates  \cite{kim}.

  In this work we show that in graphene-based anisotropic Dirac systems, as they arise in modified  (i.e. strained, or artificially
   engineered)  graphene, a rich variety of behavior
   can take place due to electron-electron interactions. 
  Within the RG approach  we find that   for  small interactions
      and anisotropy, the system flows towards the conventional isotropic fixed point.
      However 
      the anisotropy  also favors a transition towards  an  excitonic
       insulator (whose exsistance is well established in the isotropic
        case \cite{khveshchenko}), making this phase accessible  for much smaller interactions  compared to isotropic graphene.
      We show that provided the anisotropy is large enough, a quasi one-dimensional phase with dominant charge density wave order
      (anisotropic excitonic insulator) 
      can form even   for  small interactions. Thus strongly anisotropic 
   graphene-based systems can relatively  easily  experience interaction-driven transitions towards  novel  electronic phases, 
    providing a promising  avenue
   for exploration of unconventional  many-body physics.
 

\textit{Dispersion anisotropy.} We consider, for definitiveness,  graphene under uniaxial strain.  Without interactions, the Hamiltonian is \cite{pereira1}: $\hat{H}_0 = v_{x}p_{x}\hat{\sigma}_{x} + v_{y}p_{y}\hat{\sigma}_{y}$,
where $\hat{\sigma}_{x}$, $\hat{\sigma}_{y}$ are the  usual  $(2\times2)$ Pauli matrices. 
Here $v_{x} = t_{2} a_{x} \sqrt{4 \eta^2 -1}$, $v_{y} = t_{2} a_{y}$ are the velocities along the $x-$ and $y-$ direction respectively.; 
$\eta = \frac{t_1}{t_2} = \frac{t_3}{t_2} $, 
$a_{x} = a/2$ and $a_{y} = \frac{\sqrt{3}}{2}a$,  with  $a \approx \sqrt{3}\times$1.42 \AA, as shown in Fig.~\ref{fig:1}(a). 
The electronic dispersion, as shown in Fig.~\ref{fig:1}(d) is:
\begin{equation}
\varepsilon({\bf p}) = \pm\sqrt{v_x^2p_x^2 + v_y^2p_y^2}, \  \  \  \ 
 \frac{v_{y}}{v_{x}} \equiv 1+\delta.
\label{e1}
\end{equation}
and we have also  defined, and will use from now on the anisotropy parameter  $\delta$.
 We will explore the range $-1\leq \delta \leq 0$, i.e. $0\leq  v_y/v_x\leq 1$, which
 in the notation of  Fig.~\ref{fig:1}(c) would imply  strain in the armchair ($y$) direction, with  $v_y< v_x$. 
Clearly $\delta=0$  corresponds to the isotropic case, while  $\delta=-1$ is the limit of decoupled chains.
Strain in the zig-zag ($x$) direction can be easily described by the same parameter range with appropriate
 relabeling of the axes. The anisotropy  parameter $\delta$ is proportional to the strain \cite{pereira1}.
 From now on we use the effective anisotropic dispersion  (\ref{e1}) which, as mentioned earlier, is relevant
  to a variety of graphene related systems, and is not necessarily due to strain.

\textit{Electron-electron interactions and anisotropy.} The bare long-range Coulomb potential in graphene
 is given by $V(\textbf{p}) = \frac{2\pi e^2}{\kappa |\textbf{p}|}$, where $\kappa$ is the appropriate dielectric constant.  
 We consider graphene at charge neutrality where the chemical potential $\mu=0$ and $V(\textbf{p})$ is unscreened. 
 The interaction effects can be incorporated into the self-energy $\hat{\Sigma}({\textbf{p},\varepsilon})$, 
 so that the  fermion Green's function (GF) is given by: 
$\hat{G}^{-1}({\textbf{p},\varepsilon})= \varepsilon\hat{\sigma}_{0} - v_{x}p_{x}\hat{\sigma}_{x} - v_{y}p_{y}\hat{\sigma}_{y} - \hat{\Sigma}({\textbf{p},\varepsilon})$,
where $\hat{\sigma}_{0}$ is the  identity matrix.  
We work in the two-loop approximation, i.e.
$\hat{\Sigma}({\textbf{p},\varepsilon})  = \hat{\Sigma}^{(2)}(\textbf{p},\varepsilon) = i\sum_{\textbf{q}} \int^{\infty}_{-\infty} \frac{d\omega}{2\pi} V^{(2)}(\textbf{p}-\textbf{q},\varepsilon-\omega) 
\hat{G}^{(0)}(\textbf{q},\omega)
$
with the effective interaction
$V^{(2)}(\textbf{q},\omega) = V(\textbf{q},\omega) +  (V(\textbf{q},\omega))^2 
\Pi(\textbf{q},\omega)$, and $\hat{G}^{(0)}$ is the free GF. 
The physical reasons behind the use of the two-loop approximation will be discussed later 
and vertex corrections are neglected as their effect is  small, similarly to the case of isotropic
 graphene \cite{misch}.
 The dynamical polarization bubble for anisotropic Dirac fermions is easily evaluated to be:
\begin{equation}{\label{e3}}
\Pi(\textbf{q},\omega) = -\frac{N}{16 v_xv_y}
\frac{v_x^2q_x^2 + v_y^2q_y^2}{\sqrt{v_x^2q_x^2 + v_y^2q_y^2 - \omega^2}},
\end{equation}
with $N$ being the number of fermion flavors ($N=4$ for graphene.)
Using the standard decomposition
$\hat{\Sigma}^{(2)} = \varepsilon \hat{\sigma}_0 \Sigma_0 + v_{x} p_{x} \hat{\sigma}_{x}
\Sigma_{x} + v_{y} p_{y} \hat{\sigma}_{y}
\Sigma_{y}$
we obtain the dressed Green's function \cite{kotov}:
\begin{equation}
\hat{G}({\bf{p}},\varepsilon) \!=\! \frac{Z(l)}{\varepsilon\hat{\sigma}_{0} - v_{x}(l) p_{x} \hat{\sigma}_{x}
-v_{y}(l) p_{y} \hat{\sigma}_{y} }
\label{e4}
\end{equation}
where
$v_{x}(l)/v_{x} = Z(l) (1+\Sigma_{x}(l))$, $v_{y}(l)/v_{y} = Z(l) (1+\Sigma_{y}(l))$, and the quasiparticle
residue is 
$Z(l) = (1-\Sigma_0(l))^{-1} \approx 1+ \Sigma_0(l)$.  At low energies, $|{\bf{p}}|\equiv p \rightarrow 0$, all quantities 
diverge logarithmically in terms of the parameter $l=\ln(\Lambda/p)$, where $\Lambda$ is the ultraviolet cutoff.
We obtain by direct calculation the divergent contributions: 
\begin{equation}{\label{e5}}
\Sigma_{0}(l) = - \alpha_x^2\frac{N}{24(1+\delta)}\ln(\Lambda/p),
\end{equation}
\begin{equation}{\label{e6}}
\Sigma_{x,y}(l) = \left\{\frac{1}{2}\alpha_x \textrm{I}_{1,3} - \frac{N}{24(1+\delta)}\alpha_x^2\left(3-4\textrm{I}_{2,4}\right)\right\}\ln(\Lambda/p).
\end{equation}
Here $\textrm{I}_i = \textrm{I}_i(\delta), i=1,2,3,4$ are evaluated as follows. Define  $\textrm{C}(\theta,\delta)=\textrm{cos}^{2}\theta+(1+\delta)^{2}\textrm{sin}^{2}\theta$, then we have
$\textrm{I}_{n}(\delta)=\frac{1}{2\pi}\int_{0}^{2\pi}\frac{\textrm{cos}^{2}\theta\textrm{d}\theta}{[\textrm{C}(\theta,\delta)]^{n/2}}$, 
for  $n=1,2$, and  
 $\textrm{I}_{3}(\delta)=\frac{1}{2\pi}\int_{0}^{2\pi}\!\!\frac{\textrm{sin}^{2}\theta\textrm{d}\theta}{\sqrt{\textrm{C}(\theta,\delta)}}$,  $\textrm{I}_{4}(\delta)=\frac{1}{2\pi}\int_{0}^{2\pi}\!\!\frac{(1+\delta)^{2}\textrm{sin}^{2}\theta\textrm{d}\theta}
{\textrm{C}(\theta,\delta)}$.

Since $1+\delta=v_y/v_x \leq1$, it is convenient to  define the dimensionless Coulomb interaction
 coupling $\alpha_{x} = \frac{e^2}{\kappa v_x}$. From Eqs.~(\ref{e5},\ref{e6}) we obtain the RG equations
  for $\alpha_x(l)$ and the anisotropy $\delta(l)$:
\begin{equation}{\label{e7}}
\frac{d\alpha_{x}}{dl} = -\frac{\alpha_{x}^2}{2}\textrm{I}_{1}(\delta) + \frac{N\alpha_{x}^3}{6(1+\delta)} (1-\textrm{I}_{2}(\delta))
\end{equation}
\begin{equation}{\label{e8}}
\frac{d\delta}{dl} = (1+\delta)\frac{\alpha_{x}}{2}(\textrm{I}_{3}(\delta)-\textrm{I}_{1}(\delta)) - \frac{N\alpha_{x}^2}{6}(\textrm{I}_{2}(\delta)-\textrm{I}_{4}(\delta))
\end{equation}
We assume $1+ \delta \geq 0.$ It is also instructive to write these equations in the small anisotropy limit, $\delta \ll 1$:
\begin{equation}{\label{e9}}
\frac{d\alpha_{x}}{dl} = -\frac{\alpha_{x}^2}{4}\left\{1- \frac{\delta}{4} + \frac{\delta^2}{16}\right\} + \frac{\alpha_{x}^3}{3} \left\{1-\frac{\delta}{2} + \frac{\delta^2}{4}\right\} 
\end{equation}
\begin{equation}{\label{e10}}
\frac{d\delta}{dl} = -\frac{\delta\alpha_{x}}{8}\left(1-\delta^2\right) +
\frac{\delta\alpha_{x}^2}{3}\left(1- \frac{\delta}{2}\right) 
\end{equation}
In the opposite  limit  of  strong anisotropy, $\delta \approx -1$, we obtain:
\begin{equation}{\label{e11}}
\frac{d\alpha_{x}}{dl} = -\frac{\alpha_{x}^2}{\pi} + 
\frac{2\alpha_{x}^3}{3}, \ 
\frac{d\delta}{dl} = (1+\delta)\frac{\alpha_{x}}{4}\ln{\left( \frac{4}{1+\delta}\right)}
 -\frac{2\alpha_{x}^2}{3}
\end{equation}
\begin{figure}[!htbp]
\centering
\includegraphics[width=0.45\textwidth]{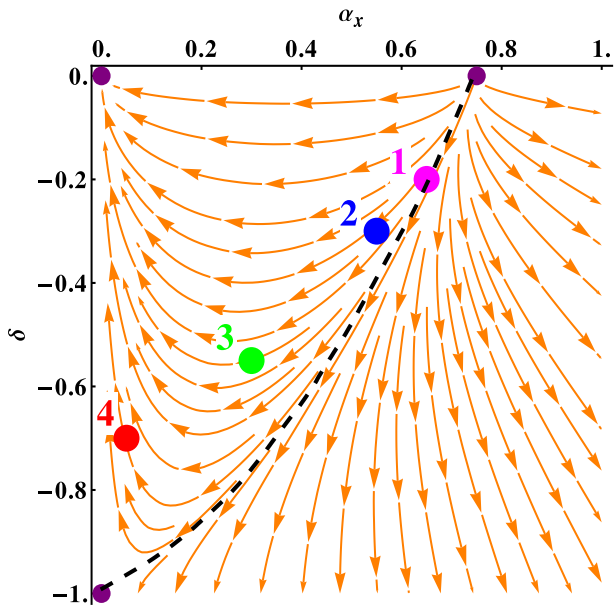}
\caption{\label{fig:2} (Color online) RG flow in the  plane $(\alpha_x,\delta)$, in terms of the interaction  
$\alpha_{x} = \frac{e^2}{\kappa v_x}$ and the anisotropy, $\delta\equiv \frac{v_y-v_x}{v_x}$. 
There is a stable fixed point at $ (0,0)$ and unstable fixed points at  $(0,-1)$ and $(3/4,0)$.  
The broken line $\alpha_{x,c}(\delta)$ separates  a region of $\alpha_x$   flowing to zero, 
for $\alpha_x < \alpha_{x,c}(\delta)$, and 
 the  region  $\alpha_x > \alpha_{x,c}(\delta)$ where $\alpha_x$  either increases  or flows towards a finite value.
The evolution of the couplings at the four points labeled by numbers are further shown in Fig.~\ref{fig:3}.}
\end{figure}

\textit{RG results: possible electronic phases.} The numerical solution  of the  RG equations, Eq.~(\ref{e7}) and (\ref{e8}), 
with $N=4$, 
leads us to the RG flow in Fig.~\ref{fig:2}  which is the main result of this work. The arrows represent the
 variations  of the couplings as the RG parameter $l=\ln(\Lambda/p)$ varies from $0$ towards $\infty$ (low-energy limit). 

First, notice that at zero anisotropy ($\delta=0$)  there is an unstable
 fixed point at $\alpha_{c} = 3/4$; the flow towards strong coupling at $\alpha_x > \alpha_{c} $ 
  corresponds to an excitonic insulator \cite{khveshchenko}, i.e.  a system with a finite gap in the spectrum 
   (see later discussion.) 
   Within the perturbative RG scheme this fixed point
   arises from the competition between the one and two loop contributions \cite{misch},
    as is clear from Eqs.~(\ref{e9}),(\ref{e10}).  
    Of course one cannot hope to reliably obtain
     the exact value of $\alpha_c $ at intermediate coupling, but the very existence of   $\alpha_c $
     has been well established by different methods \cite{khveshchenko}, with the result $\alpha_c \approx 1.1$.
     Thus we use the two-loop result in $\alpha_x$ and then take into account the anisotropy exactly.
     The dashed  line $\alpha_{x,c}(\delta) $  in Fig.~\ref{fig:2}  separates a phase with  a flow towards the stable fixed
      point at $\alpha_x=0, \delta =0$ and  a region with a flow towards large (diverging) or finite  $\alpha_x$.
      We will show  shortly that the latter case, $\alpha_x >  \alpha_{x,c}(\delta)$, is indeed characterized by a 
      divergent susceptibility towards 
    excitonic insulator. 
      For small $\delta$  we find that $\alpha_x(l)$ becomes large beyond the modified excitonic
       transition line, which can be calculated from Eq.~(\ref{e9}) to be: 
$\alpha_{x,c}(\delta) \approx (3/4)(1-|\delta|/4)$. Thus the anisotropy shifts the excitonic  transition 
to smaller coupling  ($\alpha_{x,c}(\delta)$ decreases as $|\delta|$ increases).
 
\begin{figure}[!htbp]
\centering
\includegraphics[width=0.43\textwidth]{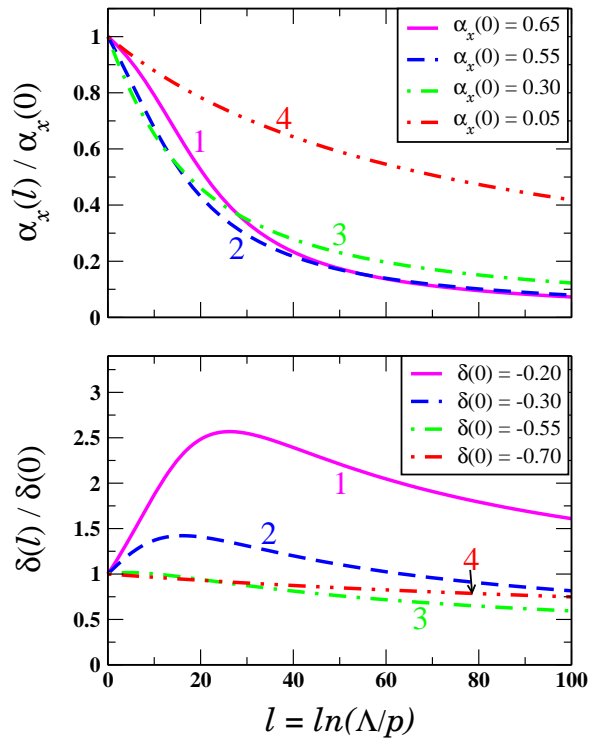}
\caption{\label{fig:3} (Color online) Evolution of the couplings $\alpha_x(l)$ (upper panel) and
$\delta(l)$ (lower panel) for initial values corresponding to the four numbered points 
from Fig.~\ref{fig:2}.}
\end{figure}

 Even below the excitonic transition line $\alpha_x <  \alpha_{x,c}(\delta) $
the anisotropy flow of $\delta(l)$ towards zero can be very slow and non-monotonic, as shown in   Fig.~\ref{fig:3}.
 This is especially true for initial values close to the excitonic line. Notice that $\delta(l)$   in fact
  increases for small $l$ up to $ l = 10 -20$.  Therefore  such a signature can be easily observed   
  for graphene at finite Fermi energy (or temperature).  For example if at $T=0$ the Fermi energy
    is $\mu \approx 0.2  {\mbox{ meV}}$  we have to  stop the RG flow at $l^{*}=\ln{(v_x\Lambda/\mu)}\approx 10$
    (taking $v_x\Lambda=5 {\mbox{ eV}}$), resulting in an increase of $\delta$ up to a factor of $2$ (or more)
     depending on the proximity to the excitonic line  (curves 1,2 in Fig.~\ref{fig:3}).

For $\alpha_x > \alpha_{x,c}(\delta)$
the region of moderate anisotropy and  relatively weak coupling, $0.4 \lesssim |\delta|$, $\alpha_x \lesssim 0.6$,
  below the dashed line   
(Fig.~\ref{fig:2}) deserves special attention. Here the flow is towards finite $\alpha_x$ and $\delta \rightarrow -1$, i.e.
 the  one-dimensional (1D) limit. From Eq.~(\ref{e11}), valid close to the 1D line, we can see that 
  $\alpha_x^{*} = 3/(2\pi) = 0.48$ separates regimes where $\alpha_x$ either experiences (relatively small)  decrease, 
   or increases under RG flow as the system  approaches the 1D  limit. Flow of $|\delta|$ beyond $1$ is not allowed
    since it reverses the sign of the velocity. 
    
\begin{figure}[!htbp]
\centering
\includegraphics[width=0.4\textwidth]{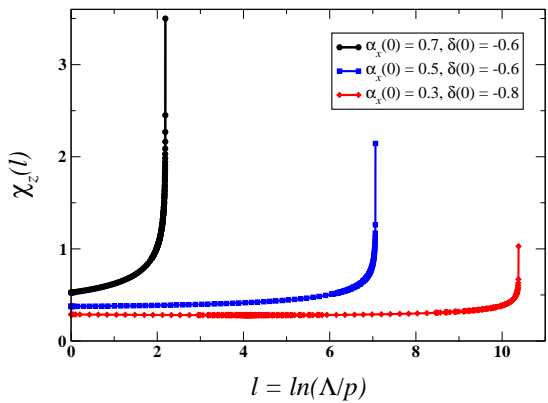}
\caption{\label{fig:4} (Color online)
Dominant divergent  susceptibility $\chi_{z}(l)$ corresponding to the order parameter 
$\langle\Psi^{\dagger}\hat{\sigma}_{z}\Psi\rangle$, for different initial couplings ($\alpha_x > \alpha_{x,c}(\delta)$). 
The divergence is at 
$l=l^{*}$ where the system approaches the 1D limit, $\delta(l) \rightarrow -1$.
}
\end{figure}

In order to characterize the nature of the phase for $\alpha_x > \alpha_{x,c}(\delta)$, we introduce
in the Hamiltonian  infinitesimal couplings $\Delta_{\mu}$
 to potential order parameters  (in the particle-hole channel): $\Delta_{\mu} \sum_{{\bf k}} \Psi_{\bf{k}}^{\dagger}\hat{\sigma}_{\mu}\Psi_{\bf{k}} $, $\mu=x,y,z$.
We then compute the interaction corrections to the corresponding vertex functions
 $\Gamma_{\mu} = \Delta_{\mu} (1+ \chi_{\mu}\ln(\Lambda/p))$ and find the 
  susceptibilities:
\begin{equation}
\label{susc}
\chi_z = \frac{\alpha_x}{2} J_z(\delta), \  \  \chi_{x,y} = \chi_z - \frac{\alpha_x}{2} J_{x,y}(\delta) .
\end{equation}
Here
$J_{z}(\delta)=\frac{1}{2\pi}\int_{0}^{2\pi}\!\!\frac{\textrm{d}\theta}{\sqrt{\textrm{C}(\theta,\delta)}}$,  
$J_{x}(\delta)=\frac{1}{2\pi}\int_{0}^{2\pi}\!\!\frac{\textrm{cos}^{2}\theta\textrm{d}\theta}
{[\textrm{C}(\theta,\delta)]^{3/2}}$, 
$J_{y}(\delta)=\frac{1}{2\pi}\int_{0}^{2\pi}\!\!\frac{(1+\delta)^{2}\textrm{sin}^{2}\theta\textrm{d}\theta}
{[\textrm{C}(\theta,\delta)]^{3/2}}$, and 
 $\textrm{C}(\theta,\delta)=\textrm{cos}^{2}\theta+(1+\delta)^{2}\textrm{sin}^{2}\theta$.
  The  quantities $\chi_{\mu}(l)$  depend on the scale $l$ via the RG running
  of  $\alpha_{x}(l), \delta(l)$. Divergence of $\chi_{\mu}(l)$  at $l=l^*$ signals spontaneous breakdown of symmetry
  in the corresponding channel \cite{andrei}. We have found that  $\chi_{z}(l)$ is the fastest divergent susceptibility 
  throughout the region $\alpha_x > \alpha_{x,c}(\delta)$, and therefore the system is an excitonic insulator
   with an   order parameter $\langle\Psi^{\dagger}\hat{\sigma}_{z}\Psi\rangle = 
   \langle\psi^{\dagger}_A\psi_A\rangle -  \langle\psi^{\dagger}_B\psi_B\rangle\neq 0$ which
    describes  charge density modulation between the two sub-lattices  \cite{khveshchenko}. 
    In Fig.~\ref{fig:4} we plot $\chi_{z}(l)$ within the perturbatively accessible parameter regime when the system
    flows towards the 1D limit, and $\alpha_x$ is finite. In this case $\chi_{z}$ diverges logarithmically:
\begin{equation}
\chi_z(l) \sim  [\alpha_{x}(l)/\pi]  \ \ln(4/[1+\delta(l)]), \ 1+\delta(l) \rightarrow 0.
\end{equation}
One can  estimate the transition temperature $T_c$ from the formula
$l^*=\ln(v_x\Lambda/T_c)$,  which gives for example (see  Fig.~\ref{fig:4}): 
  $T_c \approx 2{\mbox K}$ ($l^*\approx10.3$), $T_c \approx  45{\mbox K}$ ($l^*\approx7$),
  $T_c \approx  5\times10^3{\mbox K}$ ($l^*\approx2.2$).
  Naturally $T_c$ increases with increasing $\alpha_x$. 

From Eq.~(\ref{susc}) we find numerically that $\chi_y$ also diverges in the 1D limit, but is 
 always smaller than $\chi_z$, meaning that nematic-type (gapless) order is competing in the ground state.
 It is interesting to note that in studies of sliding Luttinger liquid (SLL)  phases \cite{sll} 
 one typically finds smectic non-Fermi liquid metals as well as charge ordered states.
  Notice that the irrelevance of the interchain hopping (our $v_y$) is one of the main characteristics of the SLL.
   Our analysis is valid, strictly speaking,  for weak coupling while non-perturbative
   methods  (such as excitonic pairing equations  as  well as bosonization) 
     are needed to further quantify the exact shape of the phase boundary in the strong coupling regime ($\alpha_x \sim 1$).
     However   the overall topology of the phase boundary as well as the quantitative behavior
      at small $\alpha_x$  are well captured within our approach.

 Finally, in the weak coupling region   $\alpha_x < 0.5$
     we find that the renormalization factor $Z$  decreases as the system approaches 1D.  From Eq.~(\ref{e5})
   we obtain  $\frac{dZ}{dl} = -\frac{Z\alpha_{x}^2}{6(1+\delta)}$, which leads to 
    the following behavior near the 1D line:
\begin{equation}{\label{e12}}
Z(l) \sim (1+\delta(l))^{1/4},\ 1+\delta(l) \rightarrow 0.
\end{equation}
Notice that $Z$ vanishes with a universal exponent $1/4$, independent of the interaction $\alpha_x$.

\textit{Summary and outlook.} In conclusion, we have found that the interplay between
the  anisotropy of the Dirac spectrum and long-range electron-electron interactions can lead to
  rich variety of behavior  and two main electronic regimes: 
  (i) a weak coupling phase characterized by a flow toward  isotropic Dirac physics  but with
   strongly renormalized parameters at low energies, and 
   (ii) a quasi one-dimensional phase with dominant charge density wave order
      (anisotropic excitonic insulator). 
    The strong anisotropic tendencies can manifest themselves in a variety of ways, such as squeezing of
     the Landau level spectrum in magnetic field \cite{goerbig}, increase of the density of states which would
      favor  itinerant ferromagnetism  and affect (increase) the specific heat \cite{kotov},  and strong anisotropies in transport
      \cite{pereira1}. The value of the interaction on typical substrates is $\alpha \approx 0.5-0.9$ \cite{kotov},
       while strain of about $15 \%$ in the zig-zag direction leads to anisotropy $|\delta| \approx 1/2$ \cite{pereira1},
        sufficient to push graphene into the anisotropic  regime.  We also  expect that artificially engineered graphene
        (optical lattices and molecular graphene \cite{optical}) can provide a promising way to investigate 
       the  strongly anisotropic states found in the present work.

\textit{Acknowledgments.} We are grateful to V.M. Pereira, E. Fradkin, B. Uchoa, F. Guinea,
M. Vozmediano, O. Vafek, H. Fertig, A. Del Maestro, and A.V. Chubukov  
for stimulating discussions and comments.
This work was supported by 
DOE grant DE-FG02-08ER46512. \\

\end{document}